\documentclass[prl,showpacs,preprintnumbers,amsmath,aps,twocolumn]{revtex4}
\usepackage{amssymb}
\usepackage{txfonts}
\usepackage{graphicx}
\usepackage{color}
\usepackage{amsmath}
\usepackage{dcolumn}
\usepackage{bm}

\newcommand{\bei}{\begin{itemize}}
\newcommand{\eei}{\end{itemize}}
\newcommand{\bee}{\begin{enumerate}}
\newcommand{\eee}{\end{enumerate}}

\begin{document}

\title{Properties of Geometric Potential in the Invariant Adiabatic Theory}
\author{Mei-sheng Zhao$^1$}
\email{zmesson@mail.ustc.edu.cn}
\author{Jian-da Wu$^1$, Jian-lan Chen$^1$ and Yong-de Zhang$^2$}
\affiliation{ $^1$ Hefei National Laboratory for Physical Sciences
at Microscale and Department of Modern Physics, University of
Science and Technology of China, Hefei 230026, People's Republic of
China.
\\ $^2$ CCAST (World Laboratory), P.O. Box 8730, Beijing 100080,
People's Republic of China}
\date{\today}

\begin{abstract}
We concentrate on the geometric potential in the invariant
perturbation theory of quantum adiabatic process which is presented
in our recent papers. It is found out to be related to the geodesic
curvature of the spherical curve in 2-dimension quantum systems. We
also show that the geometric potential
 may affect adiabatic approximation remarkably.
\end{abstract}


\maketitle


The Quantum Adiabatic Theorem is one of the most interesting
conclusions in quantum theory \cite{t1,t2,t3,t4,t5,t6,t7}. It
asserts that if the Hamiltonian of a time-dependent system varies
infinitely slowly, then the system would always remain in the state
possessing the same value of a certain dimensionless quantum number
set as the initial state. Of course, \emph{varies infinitely slow}
is only a mathematical limitation. Denote the instantaneous
eigenstates of a non-degenerate time-dependent quantum system as
$\left\{ |k\rangle,\;k=1,2,\cdots \right\} $ with corresponding
instantaneous eigenvalues $\left\{E_{k},\;k=1,2,\cdots \right\} $,
the traditional quantum adiabatic approximation condition may be
expressed as follows
\begin{equation}
\frac{|\langle m|\dot{n}\rangle |}{|E_{m}-E_{n}|}\ll 1,
\end{equation}
the dot here and below expresses the derivative with respect to
time. The validity of this condition had never been doubted until
recent years\cite{Marzlin,Tong}. They pointed out that the validity
of the traditional adiabatic approximation condition can not
guarantee the validity of adiabatic approximation. Many efforts have
been made to obtain new sufficient criterions of the adiabatic
approximation\cite{Wu,Duki,Ye,Comparat,Mackenzie,Engleman}. But none
of them achieves a complete success because the conditions given by
them are either too complicated or much more rigorous than
necessary, therefore, are inconvenient to use. Nowadays with the
development of the technique, more and more artificial
micro-structures and controllable quantum systems appears, so
time-dependent quantum systems are more and more important and
quantum adiabatic approximation is more and more interesting. In our
recent papers\cite{jianda,chen} we present an invariant perturbation
theory of quantum adiabatic proccess and proposed a new adiabatic
approximation condition according to the invariance under
time-dependent \emph{U(1)} transformation. In this paper we will
continue to study the physical and geometric meanings of the
geometric potential in our adiabatic condition.

Consider a general $d$-dimension time-dependent quantum system, let
us denote its Hamiltonian as $H(\tau),\;\tau\in\{0,T\}$. The
instantaneous eigenstates of the Hamiltonian are $\{|k(\tau)\rangle,
k=1,2,\cdots,d\}$, with corresponding energy eigenvalue
$E_{k}(\tau), k=1,2,\cdots,d\}$. Denote $ \gamma_{nm}(\tau)\equiv
i\langle n(\tau)|\dot{m}(\tau)\rangle $ , we can construct the
\emph{U(1) invariant adiabatic basis} \cite{jianda,chen}
\begin{equation}
|\Phi_{n}^{adi}\rangle=e^{-i\int_{0}^{\tau}E_{n}(\lambda)d\lambda+i\int_{0}^{\tau}\gamma_{nn}(\lambda)d\lambda}|n(\tau)\rangle.
\end{equation}
We have derived an adiabatic approximation condition\cite{jianda}
\begin{eqnarray}
\left| \frac{d}{d\tau }\arg \langle \Phi _n^{adia} (\tau ) | \dot
\Phi _m^{adia} (\tau ) \rangle \right| \gg \left| \langle \Phi
_n^{adia} (\tau ) |  \dot \Phi _m^{adia} (\tau )
\rangle \right| \nonumber \\
(\forall n \neq m ),
\end{eqnarray}
 which can be rewritten as follows
\begin{eqnarray}
&&\left| {E_n (\tau ) - E_m (\tau ) + { \gamma _{mm} \left( \tau
\right) -  \gamma _{nn} \left( \tau  \right) + \frac{d}{{d\tau
}}\arg  \gamma _{nm} \left( \tau  \right)} } \right| \gg
\left| { \gamma _{nm} \left( \tau \right)} \right|\nonumber \\
&&\quad\quad\quad\quad\quad\quad\quad\quad\quad\quad\quad\quad\quad\quad\quad\quad\left(\forall
\;n \ne m\right).
\end{eqnarray}
Compared with the traditional adiabatic approximation condition, the
new condition Eq.(4) has one extra term
\begin{eqnarray}
\Delta_{mn} &\equiv&
\gamma_{mm}(\tau)-\gamma_{nn}(\tau)+\frac{d}{d\tau}\arg
\gamma_{nm}(\tau) \nonumber
\\  &=& i\left( \langle m|\dot{m}\rangle-\langle n|\dot{n}\rangle
\right)+\frac{d}{d\tau}\arg \langle n|\dot{m}\rangle,
\end{eqnarray}
which is referred to as \emph{quantum geometric
potential}\cite{jianda,chen}.

We have revealed the invariance of this Geometric Potential under
time-dependent \emph{U(1)} transformation in \cite{jianda}. Here we
will show the relation between the geometric potential and the
geodesic curvature of spherical curve in $2$-dimension
time-dependent quantum systems. Generally, we can write the
Hamiltonian of a 2-dimension system as $H\left( \tau\right) =
A\left( \tau \right) + B\left( \tau \right) \vec{n} \left(\tau
\right) \cdot \vec \sigma , $ where $\vec{n} \left(\tau \right) =
\left( {\sin \theta \left( \tau \right)\cos \varphi \left( \tau
\right),\sin \theta \left( \tau \right)\sin \varphi \left( \tau
\right),\cos \theta \left( \tau \right)} \right)$. Choosing
appropriate phases, the Hamiltonian's instantaneous eigenstates or
\emph{adiabatic orbits} read
\begin{eqnarray}
\left\{ \begin{array}{l}
\left| { + ,\tau} \right\rangle  = \cos \frac{{\theta \left( \tau \right)}}{2}\left| 0 \right\rangle  + e^{i\varphi \left( \tau \right)} \sin \frac{{\theta \left( \tau \right)}}{2}\left| 1 \right\rangle  \\
\left| { - ,\tau} \right\rangle  = \sin \frac{{\theta \left( \tau \right)}}{2}\left| 0 \right\rangle  - e^{i\varphi \left( \tau \right)} \cos \frac{{\theta \left( \tau \right)}}{2}\left| 1 \right\rangle  \\
\end{array} \right..
\end{eqnarray}
It's quite clear that polarization vectors of the above two
adiabatic orbits point to $\vec{n} \left(\tau \right)$ and $-\vec{n}
\left(\tau \right)$ at time $\tau$, respectively. Considering the
adiabatic orbit $|+,\tau\rangle$, the geometric potential of this
orbit can be easily calculated as
 \begin{equation}\Delta _{mn}  = \frac{{\dot \theta
\ddot \phi \sin \theta + 2\dot \theta ^2 \dot \phi \cos \theta  +
\dot \phi ^3 \sin ^2 \theta \cos \theta  - \dot \phi \ddot \theta
\sin \theta }}{{\dot \theta ^2  + \left( {\dot \phi \sin \theta }
\right)^2 }}.
\end{equation}
As a comparison, we will calculate the geodesic curvature of the
spherical curve $\vec{r} \left(\tau \right) = \vec{n} \left(\tau
\right)$.
\begin{eqnarray}
\rho  &=& \left( \vec{r} \times \frac{d\vec{r}}{ds} \right) \cdot
\frac{d^2 \vec{r} }{ds^2 } \nonumber \\
&=& \frac{{\dot \theta \ddot \phi \sin \theta  + 2\dot \theta ^2
\dot \phi \cos \theta  + \dot \phi ^3 \sin ^2 \theta \cos \theta  -
\dot \phi \ddot \theta \sin \theta }}{{\left( {\sqrt {\dot \theta ^2
+ \left( {\dot \phi \sin \theta } \right)^2 } } \right)^3 }},
\end{eqnarray}
where curve element $ ds = \left| d\vec{r} \right| = \sqrt {\dot
\theta  + \left( {\dot \phi \sin \theta } \right)^2 } d\tau $. Then
we get \begin{equation}\Delta _{mn} = \rho
\frac{{ds}}{{d\tau}}.\end{equation} Same result will be obtained in
the case of adiabatic orbit $|-,t\rangle$ with corresponding
spherical curve $\vec{r}(t)=-\vec{n}(t)$. This result shows a
differential geometric property of the geometric potential. Besides,
if we integrate the geometric potential over a close smooth curve we
will obtain the difference of Berry phase between different
adiabatic orbits
\begin{eqnarray}
 \oint {\Delta _{mn} d\tau }  &=&{\arg \langle n |
 {\dot m} \rangle }\mid_0^T  + i\left( {\oint {\left\langle m \right|} \left. {\dot m} \right\rangle d\tau  - \oint {\left\langle n \right|} \left.
  {\dot n} \right\rangle d\tau } \right) \nonumber \\
  &=& i\left( {\oint {\left\langle m \right|} \left. {\dot m} \right\rangle d\tau  - \oint {\left\langle n \right|} \left. {\dot n} \right\rangle d\tau }
  \right),
\end{eqnarray}
which shows that the geometric potential also holds an integral
geometric property.

In the following part, we will present some examples to show the
significant effects caused by the geometric potential on the quantum
adiabatic approximation. Let us study a modification of the model
investigated in ref.\cite{jianda}. The Hamiltonian is given as below
\begin{equation}
H(\tau ) = \eta \sigma _z  + \xi \left[ {\sigma _x \cos \left(
{2K\eta \tau } \right) + \sigma _y \sin \left( {2K\eta \tau }
\right)} \right],
\end{equation}
where $\eta>0,\;\xi>0\;\text{and}\;K$ are all constant parameters.
For this kind of Hamiltonian Eq.(3) or Eq.(4) is a sufficient
criteria for adiabatic approximation\cite{chen}. Choosing
appropriate phases, the two adiabatic orbits can be written in
following form
\begin{eqnarray}
\left\{ \begin{array}{l}
\left| { + ,\tau } \right\rangle  = \cos \left( {\frac{\theta }{2}}
 \right)\left| 0 \right\rangle  + e^{2iK\eta \tau } \sin \left( {\frac{\theta }{2}} \right)\left| 1 \right\rangle  \\
\left| { - ,\tau } \right\rangle  = \sin \left( {\frac{\theta }{2}} \right)\left| 0 \right\rangle  - e^{2iK\eta \tau } \cos \left( {\frac{\theta }{2}} \right)\left| 1 \right\rangle  \\
\end{array} \right.,
\end{eqnarray}
where $\cos \theta  = \eta /\sqrt {\eta ^2 + \xi ^2 }$. Consider
adiabatic orbit $|+,\tau\rangle$, we can calculate the geometric
potential $\Delta _{+-}  = 2K\eta \cos \theta $. It is easy to
obtain the expression of the our adiabatic condition
\begin{equation}{\sqrt {\eta ^2  + \xi ^2 }  -
K\eta \cos \theta } \gg K\eta \sin \theta.
\end{equation}
Suppose the initial state of the system is $|+,0\rangle$, evolution
states or \emph{dynamic evolution orbit} reads
\begin{equation}
|\Psi(\tau)\rangle=e^{ - iK\eta \sigma _z \tau } e^{ - i\left(
{\left( {1 - K} \right)\eta \sigma _z  + \xi \sigma _x } \right)\tau
}\left|+,0\right.\rangle .
\end{equation}
We can calculate the
fidelity between the dynamic evolution orbit and the adiabatic orbit
at time $\tau$
\begin{eqnarray}
F\left( \tau \right)& =& \left| {\left\langle { +
,\tau } \right|U\left(
{\tau ,0} \right)\left| { + ,0} \right\rangle } \right| \nonumber \\
&=& \sqrt {\cos ^2 \left( {A\tau } \right) + \sin ^2 \left( {A\tau }
\right)\left[ {\frac{{\left( {1 - K} \right)\eta \cos \theta  + \xi
\sin \theta }}{A}} \right]^2 }, \nonumber
\end{eqnarray}
where $A =
\sqrt {\left( {1 - K} \right)^2 \eta ^2  + \xi ^2 } $ is also a
constant parameter.

If $K<0$, both the traditional adiabatic condition and our condition
guarantee the validity of the adiabatic approximation. If $K>0$,
there are two cases to which should be paid special attentions.

For the first case, we may choose $\eta\gg\xi$ and $K\simeq1$, then
the traditional condition is satisfied but our condition is not.
Meanwhile, the fidelity $F\left( \tau \right) \approx \sqrt {1 -
\cos ^2 \theta \sin ^2 \left( {A\tau } \right)} \nrightarrow 1$ when
$\tau$ is not too small. Thus, even though the traditional condition
is satisfied and we might regard the system as slowly changing one,
the quantum adiabatic approximation may be unfaithful description of
the system because of the effect of the geometric potential. Fig.1
shows both the trajectory of polarization vectors of the dynamic
evolution orbit and the adiabatic orbit on Bloch sphere surface when
$K=1$, $\eta=1$ and $\xi=0.1$. In the remain part of this paper, we
will not distinguish state and its polarization vector and call just
the former as evolution orbit and the latter as adiabatic orbit for
simplicity.

\begin{figure}[h]
\begin{center}
\includegraphics[width=0.3\textwidth]{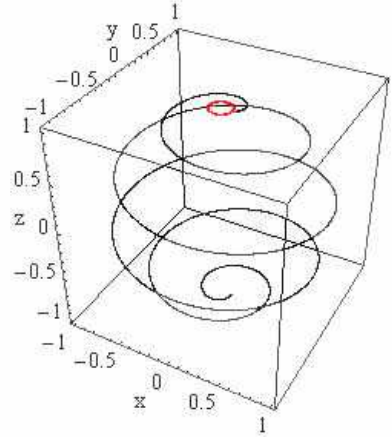}
\end{center}
\caption{evolution orbit(the black one) will be far away from
adiabatic orbit(the red one) after several cycles of Hamiltonian}
\end{figure}

For the second case, we choose $\eta\gg\xi$ but $K\gg1$. In this
case the geometric potential is much larger than the difference of
the instantaneous energy eigenvalues, and the our adiabatic
condition is satisfied while the traditional one is not. Now we have
$F\left( \tau \right) \approx \sqrt {1 - \sin ^2 \theta \sin ^2
\left( {A\tau } \right)} \approx 1$. Therefore, the geometric
potential can help to guarantee the validity of the adiabatic
approximation despite the difference of energy eigenvalues is too
small to satisfy the traditional condition. Fig.2 shows evolution
orbit and adiabatic orbit for $K=10$, $\eta=1$ and $\xi=0.1$. Fig.3
shows details after one cycle of Hamiltonian and Fig.4 shows details
after several cycles of Hamiltonian.

\begin{figure}[h]
\begin{center}
\includegraphics[width=0.3\textwidth]{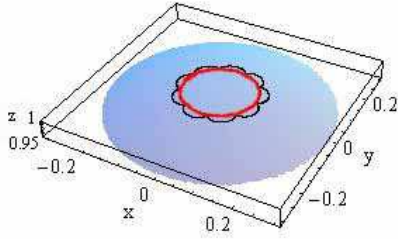}
\end{center}
\caption{evolution orbit and adiabatic orbit}
\end{figure}

\begin{figure}[h]
\begin{center}
\includegraphics[width=0.3\textwidth]{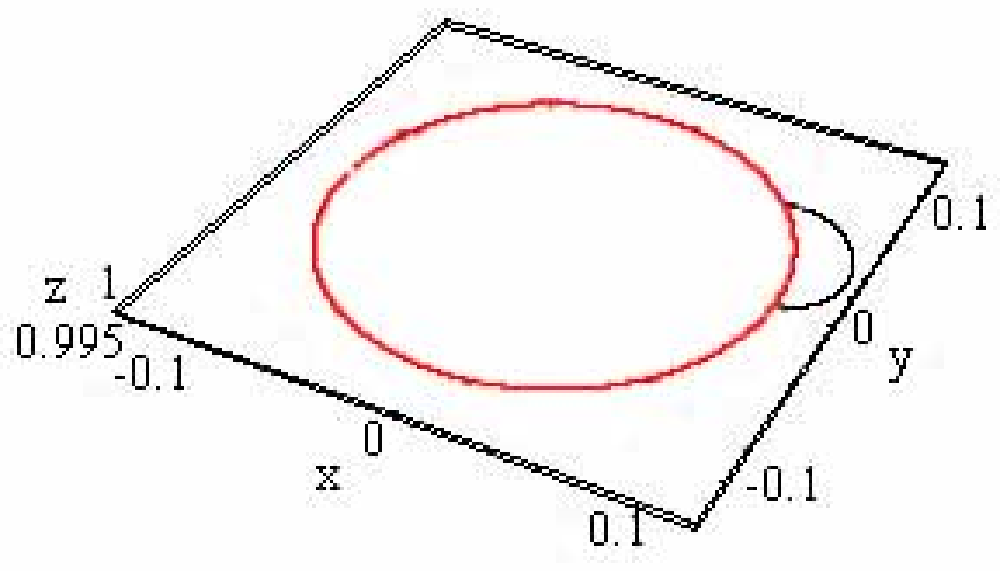}
\end{center}
\caption{Details of evolution orbit(the black line) and adiabatic
orbit(red line) after one cycle of Hamiltonian }
\end{figure}

\begin{figure}[h]
\begin{center}
\includegraphics[width=0.3\textwidth]{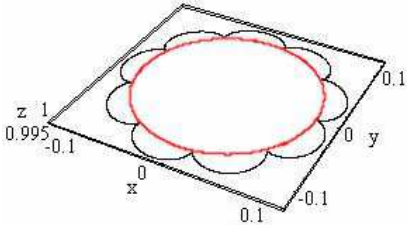}
\end{center}
\caption{Details of evolution orbit(the black line) and adiabatic
orbit(red line) after many cycles of Hamiltonian}
\end{figure}
The lower bound of the fidelity $F(\tau)$ is
$\left|\left((1-K)\eta\cos\theta+\xi\sin\theta\right)/A\right|$. If
$K\gg1$, the lower bound of fidelity can be approximated to be
$\cos\theta$. The difference of the two adiabatic orbit's Berry
phase is $2K\eta T\cos\theta=2\pi\cos\theta$, where the $T$ is the
cycle period of the Hamiltonian. So we can conclude that larger the
difference of the different adiabatic orbits' Berry phase is, more
precise the quantum adiabatic approximation will be. And the
conclusion is always correct in cases that the difference of the
system's energy eigenvalues is small and geometric potential itself
guarantees the validity of the adiabatic approximation.

At last, we will show a counterintuitive example. The Hamiltonian is
given as below
\begin{equation}\left\{
\begin{array}{l}
H=\vec{C} \left(\tau \right) \cdot \vec{\sigma}/2, \\
\vec{C} \left(\tau \right)= f \left( \tau \right)\vec{n} \left( \tau
\right) + \vec{m} \left( \tau \right)
\end{array}\right.,
\end{equation}
where $\vec{n} \left( \tau \right) = \big( {\sin \theta(\tau) \cos
\phi(\tau) ,\sin \theta(\tau) \sin \phi(\tau) ,\cos \theta(\tau) }
\big)$ and $\vec{m} \left( \tau \right) = \left( {\dot \theta(\tau)
\sin \phi(\tau) , - \dot \theta(\tau) \cos \phi(\tau) ,\dot
\phi(\tau) } \right)$ . Set the initial state be the eigenstate of
the above Hamiltonian at initial time, it is easy to find out that
if $\theta \left( 0 \right) = \dot \theta \left( 0 \right) = 0$, the
evolution orbits of the given Hamiltonian read
\begin{equation}
\left\{ \begin{array}{l}
 \left| {\psi _ +  \left( \tau \right)} \right\rangle  = e^{i\int_{0}^{\tau} f(\lambda)d \lambda /2}\left( \cos
\frac{\theta(\tau)}{2}|0\rangle+e^{i\phi(\tau)}\sin\frac{\theta(\tau)}{2}|1\rangle\right) \\
 \left| {\psi _ -  \left( \tau \right)} \right\rangle  = e^{ - i\int_{0}^{\tau} f(\lambda)d \lambda /2}\left( \sin
\frac{\theta(\tau)}{2}|0\rangle-e^{i\phi(\tau)}\cos
\frac{\theta(\tau)}{2}|1\rangle\right) \\
 \end{array} \right..
\end{equation}
We choose $\varphi \left( \tau \right) = {\rm{5\tau +
0.15sin[20\tau]}}$, $ \theta \left( \tau \right) = \tau^{1.1} /50\pi
$, and $\phi \left(\tau \right) = 0.2\tau$, evolution orbit and
adiabatic orbit on Bloch sphere surface from time $\tau=0$ to
$\tau=90\pi$ are shown in Fig.5 and Fig.6 respectively.
\begin{figure}[h]
\begin{center}
\includegraphics[width=0.3\textwidth]{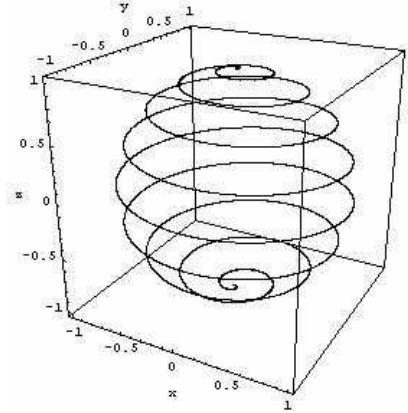}
\end{center}
\caption{Evolution orbit from $\tau=0$ to $\tau=90\pi$}
\end{figure}
\begin{figure}[h]
\begin{center}
\includegraphics[width=0.3\textwidth]{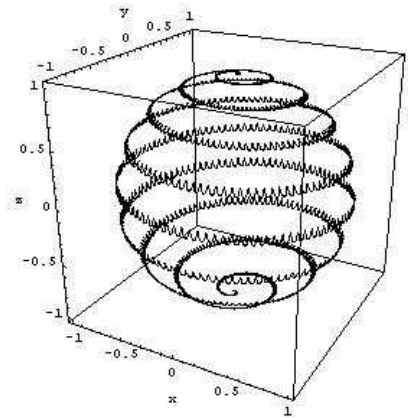}
\end{center}
\caption{Adiabatic orbit from $\tau=0$  to $\tau= 90\pi$}
\end{figure}
We can see the adiabatic orbit is a more complicated curve on Bloch
sphere than the evolution orbit while they always hold a high
fidelity $F\simeq1$ during time $\tau=0$ to $\tau=90\pi$. It may be
of some interests as it is a little different from the common
opinion about adiabatic approximation process.

In this paper we show the differential and integral geometric
properties of the geometric potential presented in our recent paper
, and then we discusses its effects on quantum adiabatic
approximation. From traditional opinion, the difference between
instantaneous energy eigenvalues $E_m \left( \tau \right) - E_n
\left( \tau \right)$ represent the time-dependent quantum system's
internal characteristic frequency. Furthermore, the existence of
geometric potential suggests that the description of the
time-dependent system's evolution might be more precise and more
appropriate if we replace the difference of the systems'
instantaneous energy eigenvalues by $E_m \left( \tau \right) - E_n
\left( \tau \right) + \Delta_{mn}$. It seems to be a pity that
$\Delta_{mn} $ does not satisfy the Rydberg-Ritz Combination
Principle(RCP) because of the existence of the term $d\arg
\left\langle {n}
 \mathrel{\left | {\vphantom {n {\dot m}}}
 \right. \kern-\nulldelimiterspace}
 {{\dot m}} \right\rangle /d\tau. $
Moreover, when the instantaneous eigenstate does not satisfy the
time-dependent Schr$\ddot{\text{o}}$dinger equation, it is not a
physical state, so RCP are not necessary to be satisfied. If
${\left\langle n \right|\left. {\dot m} \right\rangle }=0,\forall
n\neq m $, the adiabatic orbit is exactly the dynamic evolution
orbit and this orbit become physical states, RCP are satisfied
automatically. What surprises us is that $\Delta_{mn}$ in our
adiabatic condition relates closely to the geometric property of the
Hamiltonian's parametric space and the adiabatic orbits. It is quite
clear that non-trival geometric properties will more or less affect
the evolution process as long as the Hamiltonian varies with time.

We thank Prof. Si-xia Yu for illuminating discussion. This work is
supported by the NNSF of China, the CAS, and the National
Fundamental Research Program (under Grant No. 2006CB921900).

\end{document}